\documentclass{PoS}

\usepackage{wrapfig}

\title{A compact electron injector for the EIC based on plasma wakefields driven by the RHIC-EIC proton beam}

\ShortTitle{RHIC-EIC proton beam plasma wakefield acceleration}

\author{\speaker{James Chappell}\\
        University College London\\
        E-mail: \email{james.chappell.17@ucl.ac.uk}}

\author{Allen Caldwell\\
        Max Planck Institute for Physics\\
        E-mail: \email{caldwell@mpp.mpg.de}}
        
\author{Matthew Wing\\
        University College London\\
        E-mail: \email{m.wing@ucl.ac.uk}}

\abstract{Initial simulations investigating using the RHIC-EIC proton beam as the drive beam in a plasma wakefield acceleration experiment are presented. The proton beam enters the plasma and undergoes self-modulation, forming a series of microbunches. These microbunches resonantly drive electron density perturbations within the plasma, exciting a longitudinal electric field with accelerating gradients in excess of $\mathrm{GVm^{-1}}$. Injecting electrons into the resulting wakefield offers an efficient method for accelerating electron bunches for use in the proposed EIC collider.}

\FullConference{XXVII International Workshop on Deep-Inelastic Scattering and Related Subjects - DIS2019\\
		8-12 April, 2019\\
		Torino, Italy}

\begin{document}

\section{Introduction}

Plasma wakefield acceleration (PWFA) is a promising technique for the acceleration of charged particles due to its ability to provide accelerating gradients that can exceed those produced via traditional RF cavities by up to three orders of magnitude. Therefore, its suitability to provide short, cost-effective accelerators for future particle physics experiments should be explored. 

PWFA relies on the response of electrons within a plasma to the fields produced by an ultra-relativistic, high density, charged particle beam as it propagates. For a proton driver, the space-charge fields induced by the drive beam act on the plasma electrons, causing them to be initially accelerated towards the drive beam's propagation axis. Plasma ions can be assumed to be stationary due to their much larger mass and hence respond on much longer timescales than electrons. As plasma electrons are accelerated towards the region following the drive beam, a region of charge imbalance is created and co-propagates behind the drive beam. Electrons that were transversely accelerated due to the drive beam are then attracted to the axis by this region but overshoot and begin to oscillate around the propagation axis. This oscillation occurs at the plasma frequency, $\omega_p = \sqrt{n_{e}e^2 / m_e \varepsilon_0}$, where $n_e$ is the plasma electron density, $e$ the electron charge, $m_e$ the electron mass and $\varepsilon_0$ the vacuum permittivity. The large charge separations induced by these coherent oscillations create electric and magnetic fields that co-propagate behind the drive beam; the plasma wakefield. The limit of the amplitude of the fields that can be supported by the plasma depends only on the plasma density and is equal to $E_0 = m_e c \omega_p / e$ in a cold plasma. Therefore, by tuning the plasma density, the amplitude of the accelerating fields can be controlled. "Witness" electrons that are externally injected into the wakefield can become captured in its structure and are subject to accelerating gradients that can exceed $\mathrm{GVm^{-1}}$ in magnitude~\cite{awakenature}.

\section{Seeded self-modulation}

Linear plasma wakefield theory dictates that in order to efficiently drive a plasma wakefield, the drive beam length $\sigma_z$ must be on the order of, or shorter than, the inverse plasma wavenumber, $\sigma_z \sim \sqrt{2}\, k_p^{-1}$, where $k_p = \omega_p/c$~\cite{linear}. In addition to this, to avoid transverse filamentation instabilities which cause breakup of the drive beam, the transverse beam size must also satisfy the relation $\sigma_r \leq k_p^{-1}$. Proton beams that are produced at accelerator facilities typically have small transverse sizes ($\sigma_r \sim 100\,\mathrm{\mu m}$) but are long ($\sigma_z \sim 10\,\mathrm{cm}$) and therefore do not appear to be suitable as drive beams for PWFA. However, it is possible to take advantage of an intrinsic plasma response to a long drive beam known as self-modulation that acts to split it into a series of microbunches~\cite{awake1}, as demonstrated in Figure 1. 

During the self-modulation process, the initial seed transverse fields driven by the long ($\sigma_z \gg k_p^{-1}$) drive bunch couple to the beam radius evolution, inducing periodic regions of focusing and defocusing fields separated by the plasma wavelength, $\lambda_p = 2\,\pi\,c/\omega_p$~\cite{selfmodulation}. These fields act to modulate the long drive bunch into a series of microbunches, each separated by the plasma wavelength, which can resonantly drive large amplitude wakefields. By matching the plasma density to the transverse size of the proton beam such that $\sigma_r = k_p^{-1}$, large amplitude wakefields with longitudinal gradients exceeding $\mathrm{GVm^{-1}}$ can still be driven by long proton drive beams~\cite{tev}.

\begin{figure}[t]
	\centering
	\includegraphics[width=0.8\textwidth]{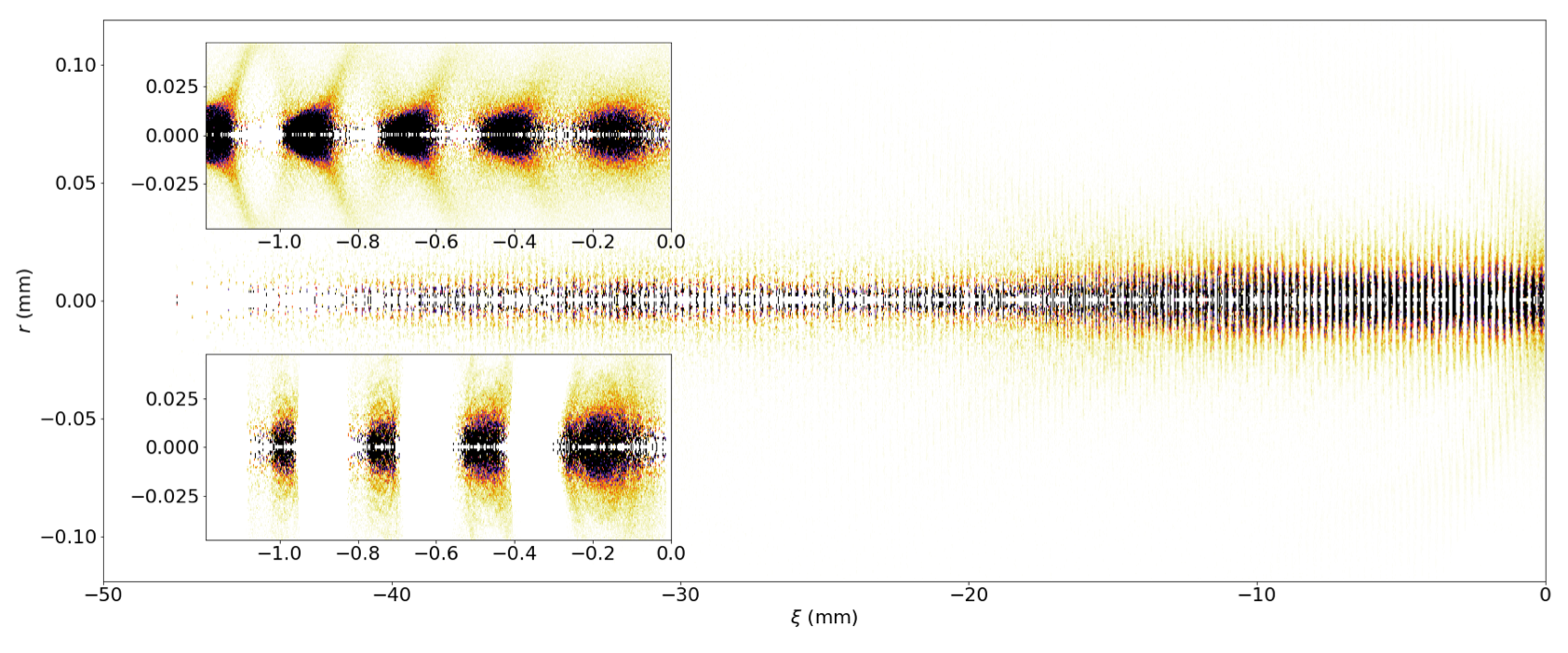}
    \caption{Example beam profile for a proton drive beam as it undergoes self-modulation. The beam parameters correspond to the "matched to ellipse" beam outlined in Table 1 at propagation distance of $z = 0.4\,$m. Inset, top left: beam profile zoomed in on the first few microbunches at $z = 0.4\,$m. Inset, bottom left: beam profile zoomed in on the first few microbunches at $z = 2\,$m, after saturation has occurred.}
    \label{fig:ssm}
\end{figure}

In order to ensure that self-modulation is the process that dominates evolution of the proton bunch, rather than competing instabilities such as hosing that can cause transverse breakup of the bunch, it is preferentially seeded by creating a sharp charge density step~\cite{hosing}. Experimentally this can be achieved by co-propagating an ionising laser pulse with the proton bunch. In this case, the plasma is created within the drive bunch, forming a sharp ionisation front which interacts with maximal bunch density, driving initial fields that are sufficiently large to seed the self-modulation process. It has been shown experimentally that this induces phase-stable modulation of the drive bunch, a requirement for consistent, repeatable witness bunch acceleration.

\section{Drive bunch parameters}

Two sets of proton bunch parameters were used within simulations to compare the wakefields that could be driven by the proposed RHIC-EIC proton beam. These are summarised in Table 1. The proton beam parameters are modified from those given in the eRHIC pre-conceptual design report when strong hadron cooling is applied. The parameters have been adjusted to correct for the asymmetry in the transverse planes in the proposed proton beam which is undesirable for driving wakefields. As such, the "matched to ellipse" case describes a proton beam with radius corresponding to a Gaussian transverse $1\sigma$ area that is equal to that of the ellipsoidal beam. The "relaxed compression" case describes a proton beam with transverse size approximately equal to the larger of the two dimensions of the ellipsoidal beam. In both cases, the plasma density $n_e$ is chosen to satisfy the linear wakefield relation $\sigma_r \cdot k_p = 1$.

\begin{table}[t]
\centering
\resizebox{0.5\textwidth}{!}{%
\begin{tabular}{ccc}
\hline
\textbf{Parameter}           & \textbf{\begin{tabular}[c]{@{}c@{}}EIC\\ Matched to ellipse\end{tabular}} & \textbf{\begin{tabular}[c]{@{}c@{}}EIC\\ Relaxed compression\end{tabular}} \\ \hline
$N_p$                        & $2\,\times\,10^{11}$                                                        & $2\,\times\,10^{11}$                                                         \\
$E_p$ (GeV)                  & 275                                                                       & 275                                                                        \\
$\sigma_r$ ($\mathrm{\mu}$m) & 40                                                                        & 100                                                                        \\
$\sigma_z$ (cm)              & 5                                                                         & 5                                                                          \\
$n_e$ (cm$^{-3}$)            & $1.8\,\times\,10^{16}$                                                      & $2.8\,\times\,10^{15}$                                                      
\end{tabular}}
\caption{Proton bunch parameters used in LCODE simulations of the wakefields driven by the proposed RHIC-EIC proton beam.}
\label{tab:bunchparams}
\end{table}

The following simulations were performed using the plasma simulation code LCODE which solves for the plasma response in a frame defined by the co-moving coordinate $\xi = z - ct$ using the quasi-static approximation~\cite{lcode}. This approximation is valid in the case where the beam evolution occurs over a longer timescale than the period of the plasma wave such that the plasma and beam evolution can be decoupled, greatly speeding up computational time. The evolution of the system used a kinetic solver for the plasma evolution including effects from ion motion. A background plasma of singly ionised rubidium was used in order to minimise effects from ion motion over the long beam length~\cite{ionmotion}. The grid size used was $\Delta \xi = \Delta r = 0.02\,k_p^{-1}$ and resolution scans were performed to ensure convergence of the solution. The longitudinal profile of the drive beam was chosen to be the decreasing half-period of a shifted cosine as this allows simulation of the interaction of a plasma with a half-cut proton beam. This would be the case in experiment when the self-modulation of the beam is seeded using a laser pulse co-propagating in the centre of the beam. The transverse beam profile is described by a Gaussian of width $\sigma_r$.

\section{Longitudinal fields}

\begin{wrapfigure}{R}{0.5\textwidth}
	\centering
	\includegraphics[width=0.5\textwidth]{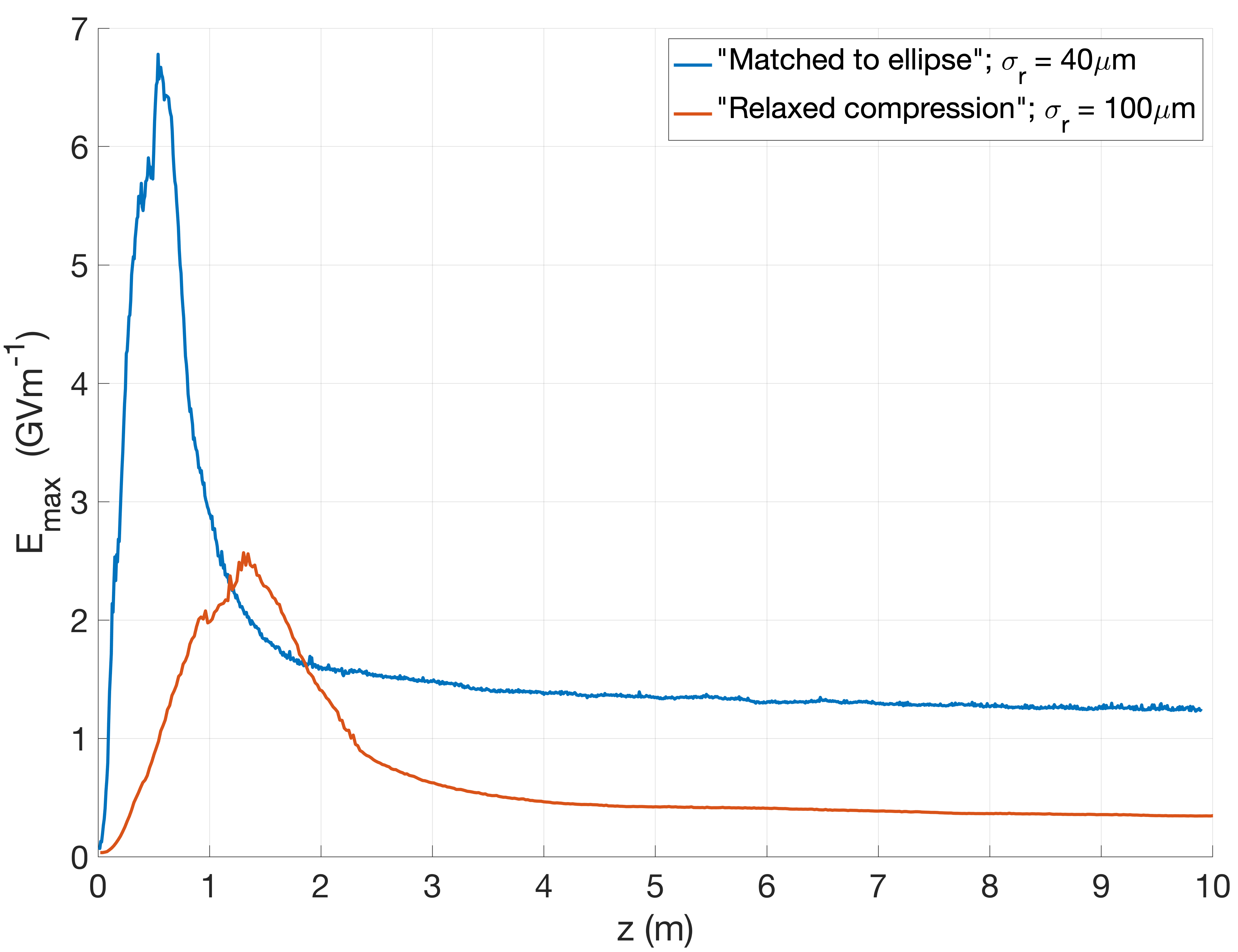}
    \caption{Evolution of the peak longitudinal fields driven by the proton drive beams over 10\,m using parameters outlined in Table 1.}
    \label{fig:long}
\end{wrapfigure}

Figure 2 shows the evolution of the maximum longitudinal fields driven by the drive beams outlined in Table 1 as the beam propagates over a distance of 10\,m. The evolution of the field strength over the first few metres is caused by the phase shift between the microbunches and the potential wells within the wakefield structure as the self-modulation process occurs~\cite{physselfmod}. Initially, the seed transverse focusing forces cause tightening of the beam in places, forming microbunches. As these continue to propagate, the potential shifts backwards with respect to the beam as the microbunches each introduce a small phase advance to the wakefields, leaving the densest part of the beam in the longitudinally decelerating fields. This is the most efficient way to drive the wakefield, causing large growth in its amplitude. The corresponding beam profile is demonstrated in the top left inset of Figure 1. As the wakefield phase continues to shift, the heads of the microbunches move into transversely defocusing regions of the wakefield and are lost from the potential well. This process shifts the microbunch centroids backwards with respect to the wakefield and thus reduces the amplitude of the potential, causing more particles to be lost from the microbunches in a positive feedback mechanism. In addition to this, the phase advance supplied by each microbunch is consequently reduced, increasing the local phase velocity of the wakefield and causing the decay in the wakefield amplitude. Eventually, a steady state is reached, known as saturation, where the bunch and wakefield phase velocities are equal, resulting in a constant amplitude wakefield. The corresponding bunch profile is shown in the bottom left inset of Figure 1. These microbunches are longitudinally shorter and have lost charge relative to those at earlier times in the self-modulation process. 

It has been demonstrated in simulation that introducing a percent-level density step within the plasma during the exponential growth period of the evolution of the wakefield freezes the elongation of the wakefield period due to the phase shift contributions from each microbunch~\cite{physselfmod}. This results in microbunches maintaining their structure over longer distances and hence larger constant amplitude fields post-saturation. Performing optimisation for these bunch parameters could further increase the energy of captured witness electrons but this has not been considered in this initial study.

\section{Witness dynamics}

\begin{wrapfigure}{R}{0.5\textwidth}
	\centering
	\includegraphics[width=0.5\textwidth]{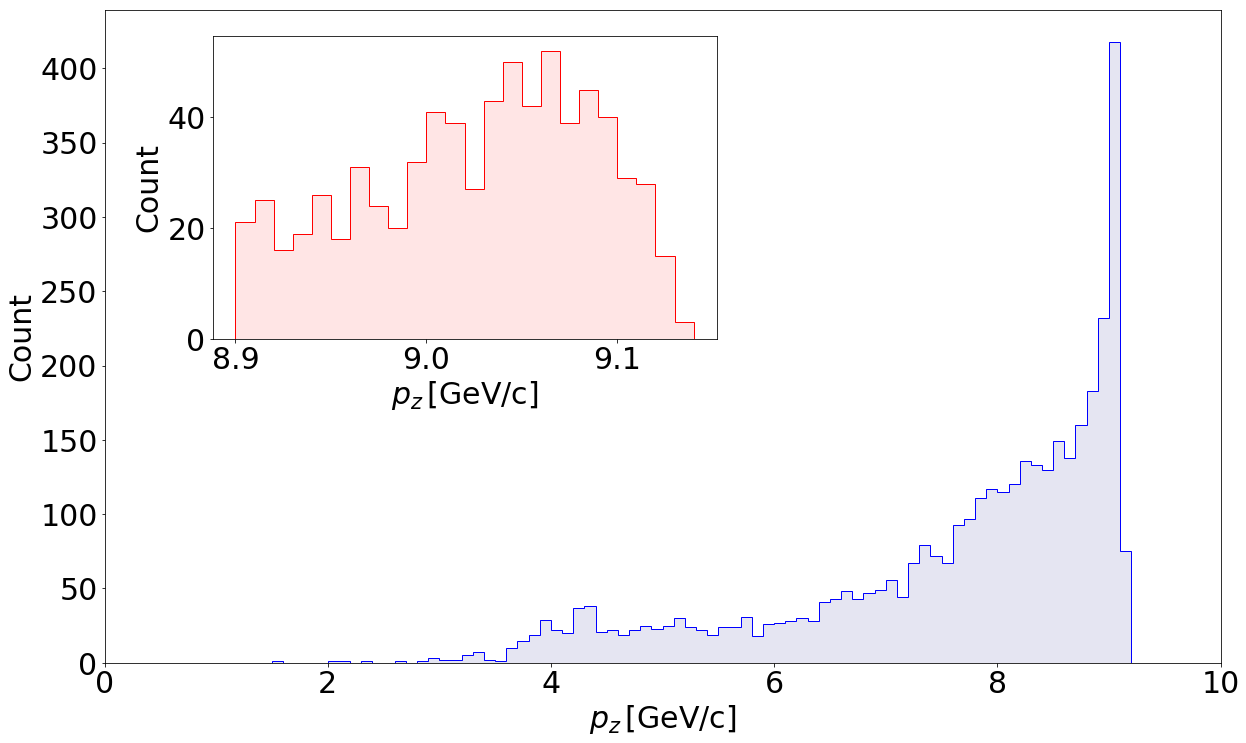}
    \caption{Example witness electron energy distribution after 8\,m of acceleration using the "matched to ellipse" drive beam parameters.}
    \label{fig:witness}
\end{wrapfigure}

The phase shift that occurs during the evolution of the self-modulation process before saturation is potentially catastrophic for acceleration of witness particles, here electrons. This is because the region of the wakefield that is both accelerating and focusing for electrons is only $\lambda_p/4$ long within a single plasma period in a phase stable wakefield. Therefore, electrons that are initially captured in this region can quickly move into a defocusing region as the wakefield phase shifts before saturation and can be lost out of the wakefield structure.

The effect of this was simulated by placing a test electron bunch within the wakefields at the start of the simulation at two $\xi$ positions behind the start of the proton bunch; $\xi$ = $-$15 and $-$30\,mm, corresponding to delays of 50 and 100\,ps respectively. Both the amplitude and rate of change of phase of the wakefields is increased for later $\xi$ positions and so finding an optimum injection position where maximum final energy is reached but wakefield phase evolution can be minimised is vital. For a delay of 50\,ps, of the 10000 test particles that were injected only 4 reached the end of the simulation at $z = 10$\,m. The majority ($>99\%$) were lost within the first 2\,m of the simulation as the self-modulation evolution occurred. For a delay of 100\,ps, where larger phase shifts are expected during self-modulation, no test electrons reached the end of the simulation. 

To overcome the limitation from the rapidly evolving phase during the self-modulation process, simulations were run where the witness electrons were instead injected after saturation had occurred and the phase-stable region had been reached. This corresponded to propagation distances of 2 and 4\,m for the "matched to ellipse" and "relaxed compression" beam parameters respectively. Experimentally, this could be realised by splitting the plasma cell into two separate stages, one for self-modulation and one for acceleration, with the electrons injected in vacuum between the two stages~\cite{awakestage}. In the "matched to ellipse" case, over 75\% of the witness electrons were captured in the wakefields and accelerated. Peak energies of the witness bunch exceeded 9\,GeV in 8\,m of plasma, demonstrating an average gradient greater than 1.1\,$\mathrm{GVm^{-1}}$. In addition to this, approximately 20\% of the accelerated electrons had energies within 1\% of this peak energy, as shown in Figure \ref{fig:witness}. The final energy distribution observed in these simulations can be further improved by matching of the witness bunch to the wakefields driven by the self-modulated proton bunch.

\section{Conclusions}

The suitability of the proposed RHIC-EIC proton beam for driving plasma wakefields using the self-modulation instability has been investigated via simulation. Peak longitudinal fields in excess of 6\,$\mathrm{GVm^{-1}}$ have been demonstrated, with average fields in excess of 1.5\,$\mathrm{GVm^{-1}}$ post-saturation. Simulations including witness electrons showed peak energies exceeding 9\,GeV in 8\,m of plasma, with approximately 20\% of captured electrons with energies within 1\% of this. This has therefore displayed the applicability of this scheme as a compact injector for the EIC machine, with the potential to provide 18\,GeV electrons in only 16\,m of accelerating medium. This acceleration length could be further reduced by introducing an optimised plasma density step that would maintain larger amplitude wakefields post-saturation.

\end{document}